# Is it possible to compare researchers with different scientific interests?

Pablo D. Batista,[1] Mônica G. Campiteli,[1]
Osame Kinouchi,[1] and Alexandre S. Martinez[1]

[1]*Faculdade de Filosofia, Ciências e Letras de Ribeirão Preto, Universidade de São Paulo*

*Avenida Bandeirantes, 3900*
*14040-901, Ribeirão Preto, SP, Brazil.*
*monicacampiteli@pg.ffclrp.usp.br*

The number $h$ of papers with at least $h$ citations has been proposed to evaluate individual's scientific research production. This index is robust in several ways but yet strongly dependent on the research field. We propose a complementary index $h_I = h / \langle N_a \rangle = h^2 / N_a^{(T)}$, with $N_a^{(T)}$ being the total number of authors in the considered $h$ papers. A researcher with index $h_I$ has $h_I$ papers with at least $h_I$ citation if he/she had published alone. We have obtained the rank plots of $h$ and $h_I$ for four Brazilian scientific communities. Contrasting to the $h$-index curve, the $h_I$ index present a perfect data collapse into a unique allowing comparison among different research areas.

## 1. INTRODUCTION

New proposals for the scientific research output evaluation have been suggested recently [1,3]. In particular, Hirsch[1,4,5,6,7] has proposed a new scalar index $h$ to quantify individual's scientific research output. A researcher with index $h$ has $h$ papers with at least $h$ citations. This index has several advantages: (i) it combines productivity with impact, (ii) the necessary data is easy to access in Thompson ISI Web of Science database, (iii) it is not sensitive to extreme values, (iv) it is hard to inflate, (v) automatically samples the most relevant papers concerning citations, etc. This index is related to extremal statistics, which is dominated by exponential density distributions, meaning that high $h$ values are difficult to achieve. Nevertheless, this index remains very sensitive to the research field. In fact, Hirsch [1] has shown that the top ten in physics and biology have very different $h$ indexes. The highest physicist (Witten E) has $h$ value equal to 110, while in the life sciences highest $h$ value (Snyder SH) is 192. Even inside a given discipline, say Theoretical and High Energy Physics, it would be hard to compare scientific research output. Further, since $h$ is an integer number, many researchers may have the same index $h$, so that discriminating or listing them becomes highly arbitrary, demanding further criteria.

To circumvent these problems it would be interesting that the $h$ index could account for the differences among disciplines. In recent papers, it has been shown that the number of citations a paper receives can be influenced by the number of authors [8]. Since: (i) the greater the number of authors, the greater the number of self-citations and (ii) the co-authorship behavior is characteristic of each discipline, we have proposed a complementary index $h_I$ to quantify an individual's scientific research output valid across disciplines [9]. The statistics of $h$ and $h_I$ are presented for the fundamental research fields in Brazil. Contrasting to $h$ rank plots, we have shown that the relative $h_I$ rank plots collapse into a single unique curve. This universal behavior suggests that it could be used to compare scientific research output performance in different research fields.

## 2. METHODOLOGY

From Thomson ISI Web of Science database, we have considered the Brazilian scientific research output in four different fields: Physics, Chemistry, Biology/Biomedical and Mathematics. The database has been compiled from the database of the Institute for Scientific Information (ISI). The search has been conducted using the query "Brazil OR Brasil" in the address field. This means that it has been accounted all the documents with at least one Brazilian address with citations till June 2005. Researcher nationality and researches done by Brazilians abroad (foreigner address) are disregarded in the considered database. We have considered all documents published from 1945 to 2004. The search has been performed separately for each year.

We have chosen the Brazilian institutions for this work due to the fact that the ISI Web of Science limits the searching to 100,000 papers, being thus impossible to compile a complete database for countries that have a greater annual productivity as the United States of America. Our database contains information of about 188,909 bibliographical references. This information includes type of publication, full reference, citations received, authors' names and addresses, including the institutions, cities, states and country. Among all publications, we have considered 150,323 articles, 24,164 meeting abstracts, 5,541 notes, 3,577 letters and 2,333 reviews. Documents have been classified into the research fields using the *tag subject*. Then four lists have been compiled containing author name, publication number, times cited and number of authors. Notice that a given researcher can appear in more than one list.

## 3. RESULTS

Figure 1 shows the number $N$ of researchers with index $h$ for the four different disciplines. The $N(h)$ distributions for different fields are apparently exponentials (not fitted). Notice however, that in Physics there exist many more researchers with $h > 10$ than the other research areas, making it more power-law like. The research fields Chemistry and Biology behave similarly.

Our data set shows that, in general biologists/biomedical researchers have smaller $h$ than physicists, in contrast to Hirsch's observations about worldwide data [1]. This may be due to the lack of financial support to sustain the experimental nature of Biology. Further, in Physics, computer and theoretical physics, which are less demanding areas, may play an important role in the mentioned trend.

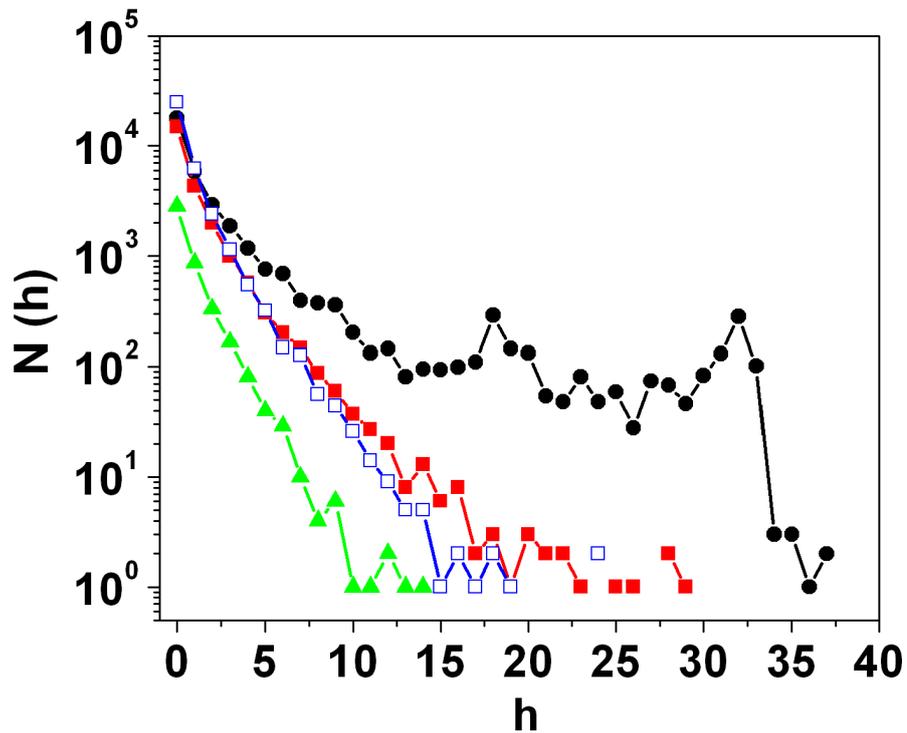

**Figure 1**. Number of researchers with $h$ index in four different research fields (● Physics, ■ Chemistry, ☐ Biology/Biomedical and ▲ Mathematics) in Brazil.

The distributions of papers with $k$ authors are shown in Fig. 2. One sees that the maximum of the distributions is at $k_{max} = 2$ for physics, biology and mathematics, being $k_{max} = 3$ for chemistry. Nevertheless, Physics have several papers with more than 50 authors. These papers probably reflect collaborations with large international teams [10]. Notice also that Mathematics presents the greater proportion of a single author papers.

We have verified that citation distributions can be fitted either by known curves, as previously reported [11, 12] (inset of Fig. 2). Also, we have empirically verified that the total number of citations ($Nc_{tot}$) approaches $ah^2$ (not shown), as conjectured in Ref [1].

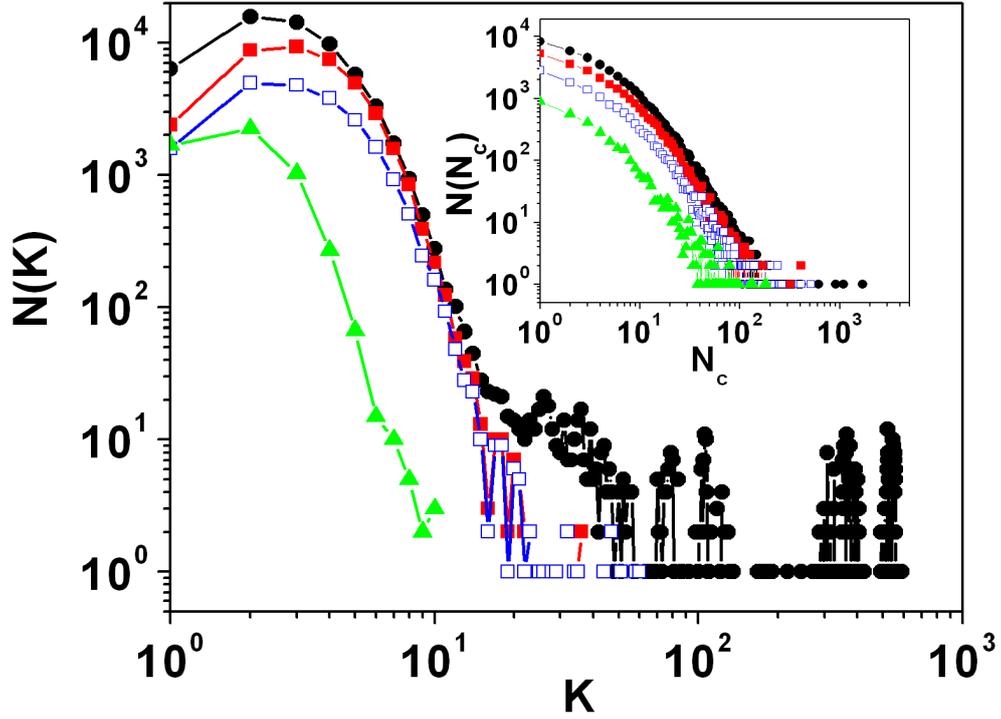

**Figure 2**. Number of publication with *k* authors per article in four research fields (● Physics, ■ Chemistry, □ Biology/Biomedical and ▲ Mathematics) in Brazil. **Inset:** Number $N(N_C)$ of publications cited $N_C$ times in four research fields in Brazil from 1945 to 2004 (citations collected till June 2005).

The top *h*-researchers in out data set are displayed in Table I. From this Table, one sees that it is very difficult to compare researchers from different fields. However, we have noticed a strong correlation between *h* and the number of authors that sign the top *h* publications.

| PHYSICS | | | | | | CHEMISTRY | | | | | |
|---|---|---|---|---|---|---|---|---|---|---|---|
| NAME | University | $h$ | $N_c$ | $N_a$ | $N$ | NAME | University | $h$ | $N_c$ | $N_a$ | $N$ |
| Eppley, G | * | 37 | 4938 | 13172 | 163 | Zagatto, EAG | USP | 29 | 2770 | 143 | 116 |
| Fisyak, Y | * | 37 | 4732 | 13218 | 156 | Toma, HE | USP | 28 | 2869 | 70 | 173 |
| Read, AL | * | 36 | 5794 | 17095 | 230 | Krug, FJ | USP | 28 | 1936 | 139 | 62 |
| Tsallis, C | CBPF | 35 | 5946 | 83 | 219 | Reis, BF | USP | 26 | 2257 | 132 | 110 |
| Yang, J | * | 35 | 3956 | 12458 | 83 | Comasseto, JV | USP | 25 | 2095 | 74 | 101 |
| Yepes, P | * | 35 | 3677 | 13415 | 120 | Dupont, J | UFRGS | 23 | 2378 | 111 | 76 |
| Alves, GA | CBPF | 34 | 3812 | 11107 | 136 | Airoldi, C | Unicamp | 22 | 2093 | 50 | 220 |
| Verbeure, F | * | 34 | 5153 | 13940 | 273 | Chaimovich, H | USP | 22 | 1642 | 102 | 68 |
| Smirnov, N | * | 34 | 4394 | 15365 | 174 | Bergamin, H | USP | 21 | 1508 | 105 | 34 |
| over 100 | | 33 | | | | Gushikem, Y | Unicamp | 21 | 1339 | 64 | 120 |
| almost 300 | | 32 | | | | Castellano, EE | USP | 20 | 1338 | 128 | 171 |
| | | | | | | Eberlin, MN | Unicamp | 20 | 1255 | 87 | 108 |
| | | | | | | Martins, MAP | UFSM | 20 | 1006 | 107 | 68 |
| | | | | | | Kubota, LT | Unicamp | 19 | 1159 | 64 | 112 |
| BIOLOGICAL / BIOMEDICAL | | | | | | MATHEMATICS | | | | | |
| deSouza, W | UFRJ | 24 | 2134 | 87 | 157 | Mane, R. | IMPA | 14 | 509 | 19 | 21 |
| Gottlieb, OR | UFF | 24 | 2657 | 87 | 222 | Iusem, AN | IMPA | 13 | 471 | 28 | 48 |
| Dobereiner, J | UFRRJ | 19 | 867 | 90 | 34 | Martinez, JM | Unicamp | 12 | 495 | 30 | 74 |
| Salzano, FM | UFRGS | 18 | 874 | 107 | 82 | Defigueiredo, DG | Unicamp | 12 | 516 | 24 | 25 |
| Arruda, P | Unicamp | 18 | 820 | 85 | 52 | Simis, A | UFPE | 11 | 393 | 29 | 39 |
| Vercesi, AE | Unicamp | 17 | 1077 | 69 | 74 | Dajczer, M | IMPA | 10 | 317 | 22 | 56 |
| Laurance, WF | INPA | 16 | 820 | 124 | 37 | Palis, J | IMPA | 9 | 230 | 17 | 19 |
| Jones, RN | Unifesp | 16 | 969 | 98 | 44 | Costa, DG | UNB | 9 | 193 | 18 | 20 |
| Yoshida, M | USP | 15 | 737 | 60 | 71 | Svaiter, BF | IMPA | 9 | 259 | 21 | 34 |
| Oliveira, OS | Unicamp | 14 | 385 | 34 | 25 | Garcia, A | IMPA | 9 | 298 | 20 | 37 |
| Curi, R. | USP | 14 | 731 | 65 | 128 | Vasconcelos WV | * | 9 | 296 | 25 | 16 |
| Sader, H S | * | 14 | 825 | 97 | 41 | Telles, JCF | UFSC | 9 | 311 | 21 | 22 |
| Trabulsi, LR | USP | 14 | 787 | 68 | 83 | | | | | | |
| Junqueira LCU | USP | 14 | 1148 | 52 | 32 | | | | | | |
| Graeff FG | USP | 13 | 556 | 36 | 46 | | | | | | |

**Table I**: The top *h*-ranking for the four fields. The numbers are (from left to right) the *h* index, total number of citations in the *h* papers, number of authors in the *h* papers and total number of papers published by the author. Authors marked with (*) are associated to foreign institutions but appear in the list because of the Brazilian collaboration in the *h* papers.

To account for the coauthorship effect, divide *h* by the mean number of researchers in the *h* publications $\langle N_a \rangle = N_a^{(T)}/h$, where $N_a^{(T)}$ is the total number of authors (author multiple occurrences are allowed) in the considered *h* papers. Thus, we obtain a new index:

$$h_I = h/\langle N_a \rangle = h^2/N_a^{(T)} \qquad (1)$$

which gives further information about the research output.

The rationale for this procedure is that we want to measure the effective individual average productivity. More authors could produce more future self-citations which may produce statistical biases. If a given researcher is the only author in his/her $h$ papers, then $N_a^{(T)} = h$ and $h_I = h$ in this case. The $h_I$ index indicates the number of papers a researcher would have written alone along his/her carrier with at least $h_I$ citations. Once $h$ has been computed, the $h_I$ index is also easy to compute from the Thompson ISI Web of Science. The rank plots of $h$ (inset of Fig. 3) and $h_I$ (Fig. 3) are strongly different.

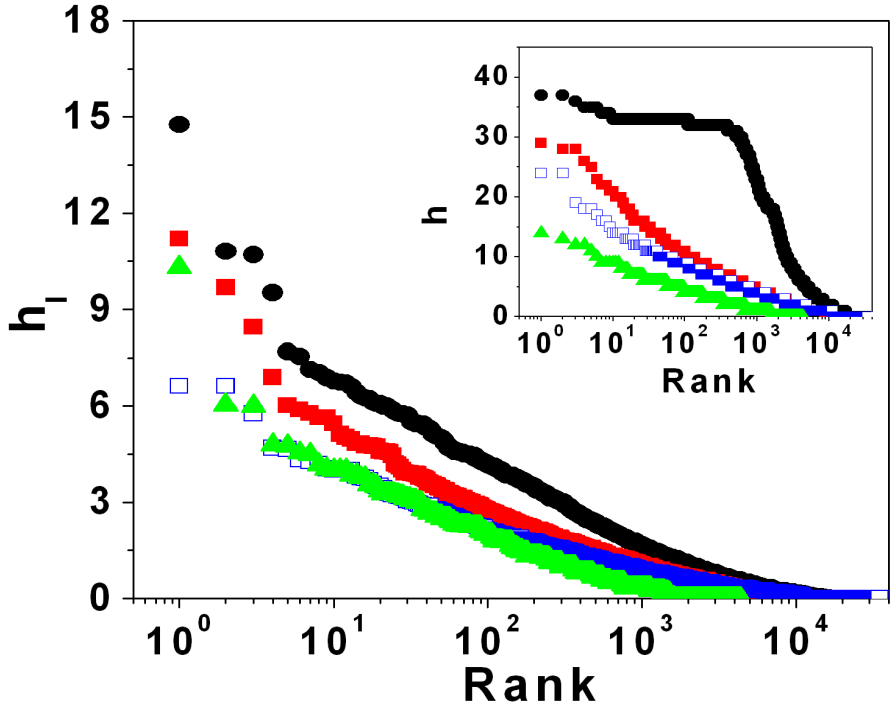

**FIG. 3:** The index $h_I$ as a function of the ranking $R$ for the brazilian research fields (● Physics, ■ Chemistry, □ Biology/Biomedical and ▲ Mathematics). The $h_I$ curves, in contrast to $h$ curves, have the same functional shape. **Inset**: The same for the index $h$.

Figure 3 presents the $h_I$ indices in a decreasing ranking plotted against the respective number of realizations. Physics rank plot is practically constant for the first 1000 $h$-ranks, presenting an abrupt decay afterwards. This rank plot drastically differs from the rank plot of other considered fields. The $h_I$ rank plot is much smoother and, importantly, all the distributions are more similar among themselves, being close to stretched exponentials (straight line in the linear-log plot) [13]. This similarity displays the emergence of a universal behavior.

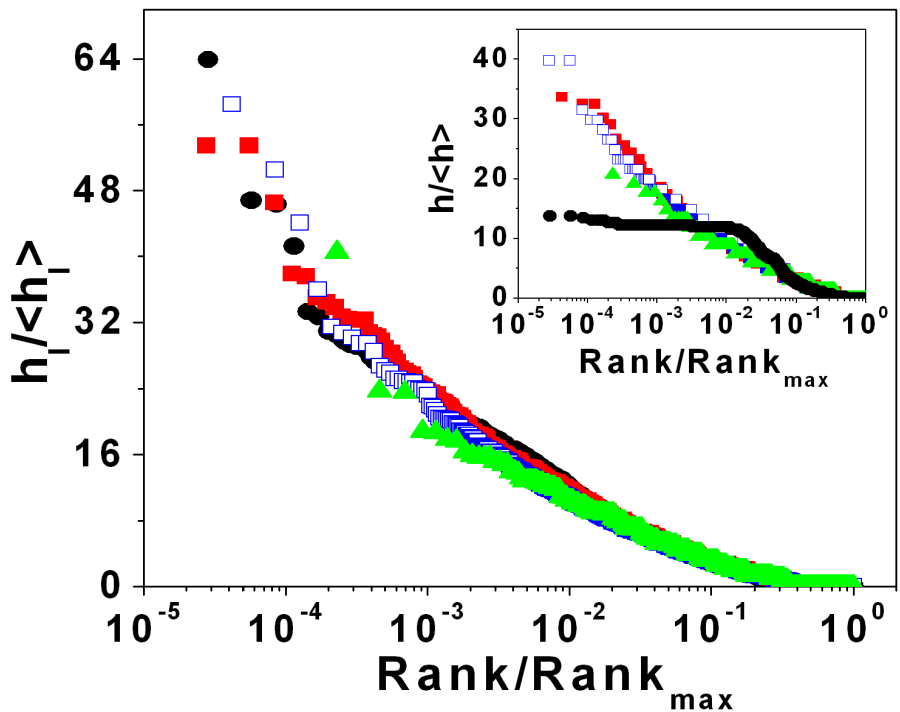

**FIG. 4:** The index $h_I/\langle h_I \rangle$ as a function of the ranking $R/R\max$ in four different research fields (● Physics, ■ Chemistry, □ Biology/Biomedical and ▲ Mathematics) in Brazil. A single unique curve is found permitting comparisons among different research fields. **Inset:** Data collapse is not obtained for $h$ curves because of the co-authorship effects in Physics.

The functional similarity of the $h_I$ rank plots has motivated us to scale the variables. With this aim, we have divided each $h_I$ curve by its mean values and the ranks have been divided by the size of the community (maximum rank). The scaled variables are plotted in Fig. 4, where the data collapse is shown by a single unique curve. This universal curve is not observed for the relative $h$ index (inset - Fig.4) since the co-authorship effects exclude Physics.

The use of the mean value in the definition of $h_I$ index could penalize authors with eventual papers with large number of authors, since the mean is a measure very sensitive to extremum values. A possible correction to this factor is to consider the median or harmonic mean instead of the mean value. In fact, we have observed a strong correlation ($r = 0.93$) between the rankings using the mean value and median measures.

| PHYSICS | | | | | | CHEMISTRY | | | | | |
|---|---|---|---|---|---|---|---|---|---|---|---|
| NAME | University | $h_I$ | $h$ | $N_c$ | $N_a$ $N$ | NAME | University | $h_I$ | $h$ | $N_c$ | $N_a$ $N$ |
| Tsallis, C | CBPF | 14.8 | 35 | 5946 | 82 219 | Toma, HE | USP | 11.2 | 28 | 869 | 69 173 |
| Berkovits, N | Unesp | 10.8 | 20 | 1101 | 36 57 | Airoldi, C | Unicamp | 9.7 | 22 | 93 | 49 220 |
| Letelier, PS | Unicamp | 10.7 | 17 | 1156 | 26 113 | Comasseto, JV | USP | 8.4 | 25 | 95 | 73 101 |
| Adhikari, S K | Unesp | 9.5 | 20 | 1423 | 41 182 | Gushikem, Y | Unicamp | 6.9 | 21 | 339 | 63 120 |
| Alcaraz, FC | USP | 7.7 | 20 | 1060 | 51 74 | Petragnani, N | USP | 6.0 | 17 | 330 | 47 50 |
| Lemos, JPS | UFRJ | 7.5 | 14 | 548 | 25 32 | Zagatto, EAG | USP | 5.9 | 29 | 770 | 142 116 |
| Nunes, OAC | UNB | 7.1 | 10 | 338 | 13 81 | Riveros, JM | USP | 5.8 | 15 | 70 | 38 57 |
| Hipolito, O | USP | 7.0 | 18 | 919 | 45 75 | Kubota, LT | Unicamp | 5.6 | 19 | 159 | 63 112 |
| Sarmento, EF | UFAL | 6.8 | 20 | 996 | 58 56 | Krug, FJ | USP | 5.6 | 28 | 936 | 138 62 |
| Swieca JA | UFSCar | 6.7 | 16 | 714 | 37 20 | Fatibello, O | UFSCar | 5.4 | 13 | 46 | 30 69 |
| BIOLOGICAL / BIOMEDICAL | | | | | | MATHEMATICS | | | | | |
| Gottlieb, OR | UFF | 6.6 | 24 | 2657 | 86 222 | Mane, R | IMPA | 10.3 | 14 | 509 | 18 21 |
| deSouza, W | UFRJ | 6.6 | 24 | 2134 | 86 157 | Iusem, AN | IMPA | 6.04 | 13 | 471 | 27 48 |
| Oliveira, PS | Unicamp | 5.8 | 14 | 385 | 33 25 | Defigueiredo D | Unicamp | 6.0 | 12 | 516 | 23 25 |
| Graeff, FG | USP | 4.7 | 13 | 556 | 35 46 | Martinez, JM | Unicamp | 4.80 | 12 | 495 | 29 74 |
| Mello MLS | Unicamp | 4.6 | 11 | 406 | 25 77 | Palis, J | IMPA | 4.76 | 9 | 230 | 16 19 |
| Ferreira, SH | USP | 4.33 | 13 | 1017 | 38 70 | Dajczer, M | IMPA | 4.54 | 10 | 317 | 21 56 |
| Peres, CA | USP | 4.27 | 8 | 254 | 14 14 | Costa, DG | UNB | 4.50 | 9 | 193 | 17 20 |
| Vercesi, AE | Unicamp | 4.2 | 17 | 1077 | 68 74 | Simis, A | UFPE | 4.2 | 11 | 393 | 28 39 |
| Lacazvieira, F | USP | 4.05 | 9 | 133 | 19 34 | Garcia, A | IMPA | 4.05 | 9 | 298 | 19 37 |
| Dobereiner, J | UFRRJ | 4.01 | 19 | 867 | 89 34 | Gonzaga, CC | UFSC | 4 | 6 | 188 | 8 15 |

**Table II:** The top $h_I$-ranking for the four fields. The numbers are (from left to right) the $h_I$ index, $h$ index, total number of citations in the $h$ papers, number of authors and total number of papers published by the author.

The top ten $h_I$-researchers in the Brazilian database are shown in Table II. The overlaps between the $h$ and $h_I$ lists are: 10% for Physics, 60% for Chemistry, 50% for Biology and 90% for Mathematics.

## 4. CONCLUSION

The index $h_I$ is complementary to $h$ and indicates the number of papers a researcher would have written along his/her carrier with at least $h_I$ citations if he/she has worked alone. It diminishes the $h$ degenerescency and has the advantage of being less sensitive to different research fields. This allows a less biased comparison due to the consideration of co-authorship effects. The $h$ ranking studied takes into account publications that have at least one author with Brazilian address and presented strong differences in functional form between fields say, Physics and Mathematics. Such differences are softened for $h_I$, where data colapse has been found with the appropriate scaling. This universal behavior allows comparisons among different fields. It may be interesting to perform this study for other countries and other instances as department evaluations, periodic publications etc.

**Acknowledge**: This work was supported by CAPES, CNPq and FAPESP.